\documentclass{ws-procs9x6}

\usepackage{epsfig,psfig}
\usepackage{graphicx}

\begin{document}

\title{Factorization and Universality \\ in Azimuthal Asymmetries\footnote{
Presented under the title: New Observables from Color Gauge Invariance}}

\author{F.~Pijlman}

\address{Department of Physics, Vrije Universiteit Amsterdam, \\
De Boelelaan 1081, NL-1081 HV Amsterdam, The Netherlands\\
E-mail: fetze@nat.vu.nl}

\maketitle

\abstracts{
The theoretical status on transverse momentum dependent
factorization in semi-inclusive DIS
and Drell-Yan is not clear
in contrast to claims in the literature in which gauge links
only at one loop were considered explicitly. Recently obtained results
beyond this order question the validity of these claims and will
be briefly discussed. Possible input from experiments to solve these
matters will be outlined.}

\section{The present theoretical status}

Single spin asymmetries (SSA's)
are full of surprises and give rise to questions
about factorization. In a factorized description of SSA's one needs a special
class, called T-odd, of \emph{transverse momentum dependent}
distribution functions or fragmentation functions.
Three separate
mechanisms were suggested to generate T-odd functions. The first
mechanism\cite{Qiu} consists of nonzero gluon fields at infinity
and was unified\cite{Boer} with the second mechanism which is
based on fully
connected gauge links\cite{Belitsky}. The non-trivial paths of these
links connect the two quark fields in the distribution and
fragmentation functions and could lead to
SSA's.
The third
mechanism, appearing only for
fragmentation functions, comes from
final state interactions\cite{Collins93}.

In a factorized description the gauge links have particular implications:
T-odd distribution functions in semi-inclusive DIS (SIDIS)
enter with a different sign in
Drell-Yan (DY)\cite{CollinsSign}; T-odd
distribution functions involve the gluon field in the
nucleon\cite{Boer}; gauge links violate naive Lorentz-invariance
relations\cite{Metz01}; links can give rise to new
functions\cite{Bacchetta}; and, links might
imply non-universal fragmentation functions\cite{Boer}. This
questions whether a factorized description
is allowed for transverse momentum dependent
observables. 

Several articles deal with factorization in SIDIS
and DY. Recently a spin dependent factorization theorem for
SIDIS and DY was claimed where the
all-order argument is based
on generalized Ward identities\cite{ji1}.
Subsequently, part of the relevant all-order
calculations were presented\cite{Bomhof}
and possible problems on factorization in DY were pointed
out\cite{Pijlman}.
Afterwards, it was suggested that the fragmentation functions and
soft factor in the factorization theorem should
contain different gauge links and the authors\cite{CollinsMetz}
claimed
universality
of distribution, fragmentation functions and
soft factors in
SIDIS, $e^+ e^-$ annihilation and DY.
The proof of universal fragmentation
functions was
formalized at one loop, but again the all-order arguments are
based on Ward identities and an explicit proof was not given.

Despite the claims on factorization
and the significant progress made in the previous references,
the present situation is not clear. However,
factorization remains essential
for comparing experimental results
and relating them to theoretical predictions.
In the next section
gauge link derivations\cite{Bomhof}
and their consequences for
factorization\cite{Pijlman} will be briefly
presented. As will be shown, Ward identities should be applied carefully.
The last section will discuss
how experiments and theory could contribute to solve these\ \nolinebreak
matters.

\section{Gauge links and factorization}

\begin{figure}[t]
\begin{center}
\begin{tabular}{ccccc}
A &\includegraphics[width=2cm]{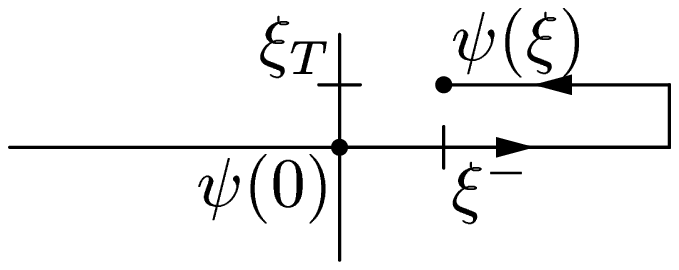}&
\includegraphics[width=2cm]{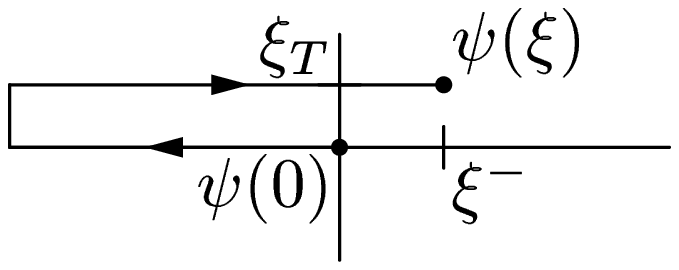}  &
\includegraphics[width=2cm]{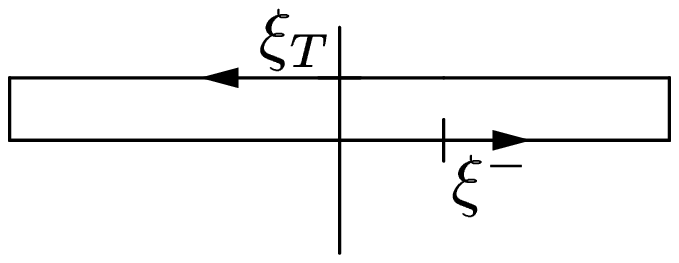}  &\\
& $\mathcal{L}^{[+]}$ & $\mathcal{L}^{[-]}$ & $\mathcal{L}^{[ \Box ]}$ & \\
B &
\includegraphics[width=2cm]{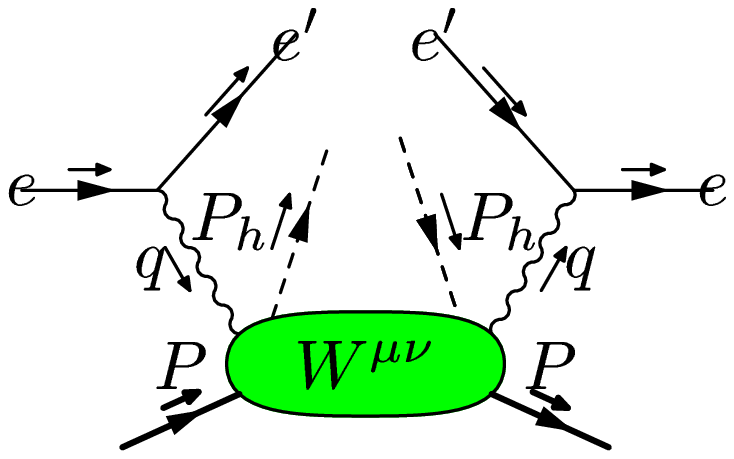}  &
\includegraphics[width=2cm]{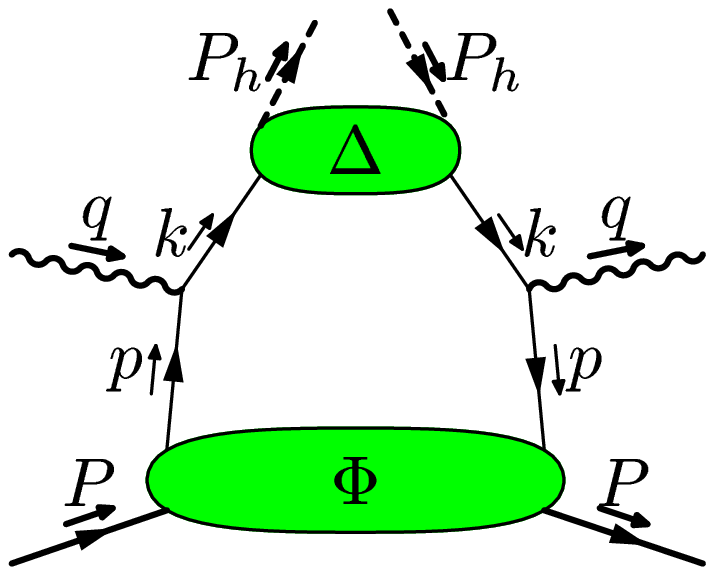} &
\includegraphics[width=2cm]{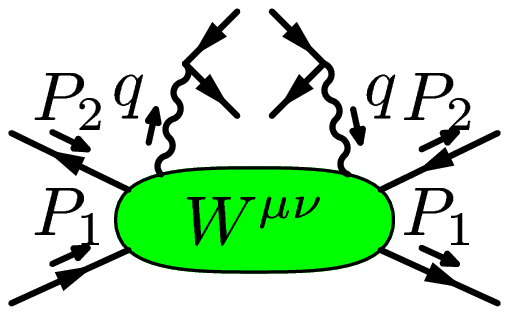} &
\includegraphics[width=2cm]{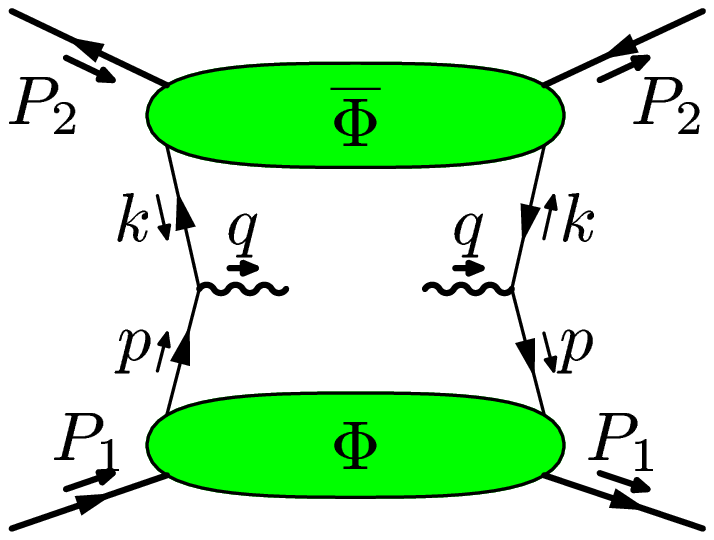}\\
& SIDIS & $\mathcal{L}^{[+]}$ & DY & $\mathcal{L}^{[-]}$\\
C &
\includegraphics[width=2cm]{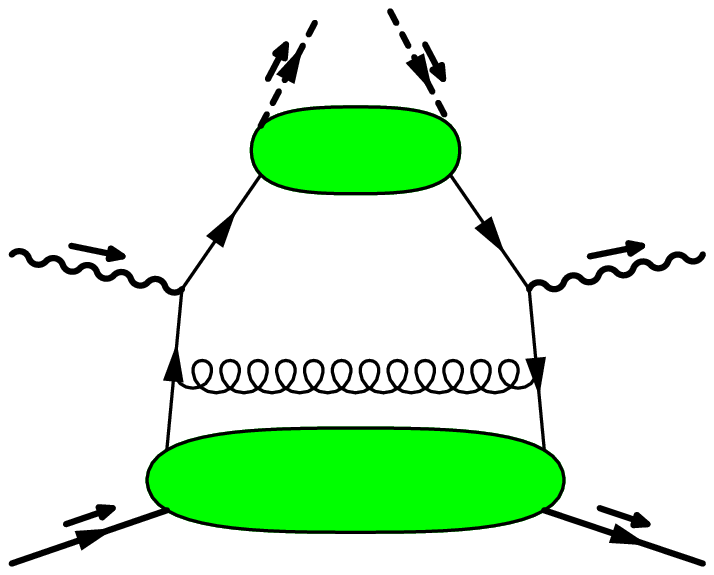}  &
\includegraphics[width=2cm]{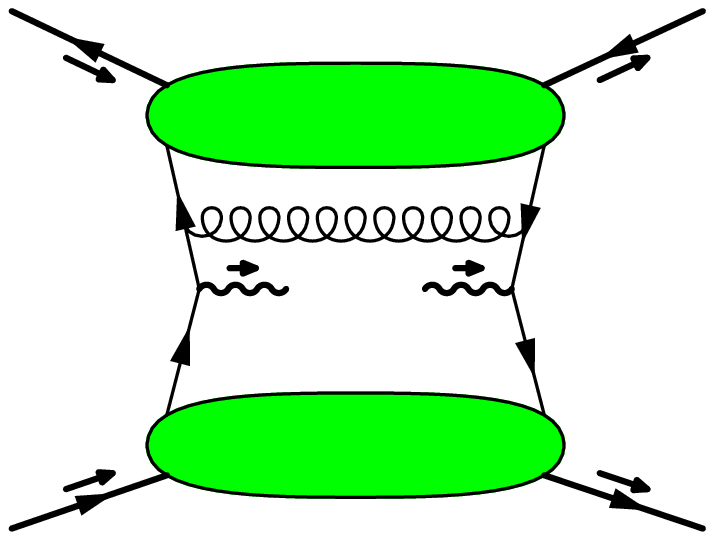}   &
\includegraphics[width=2cm]{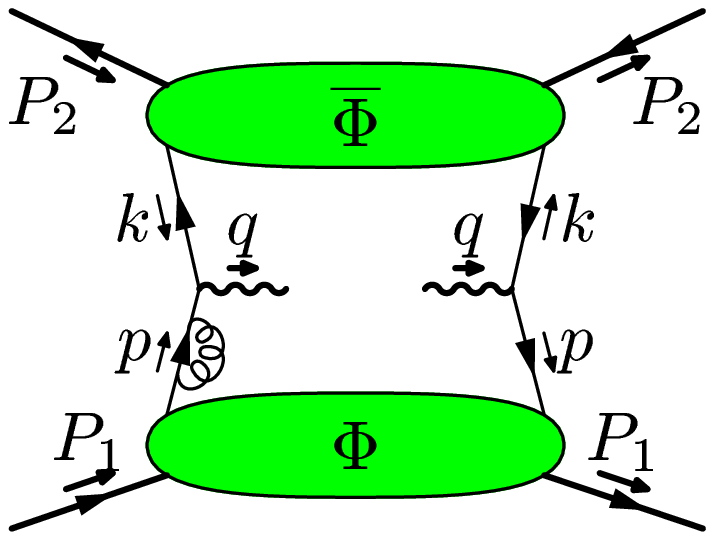} &
\includegraphics[width=2cm]{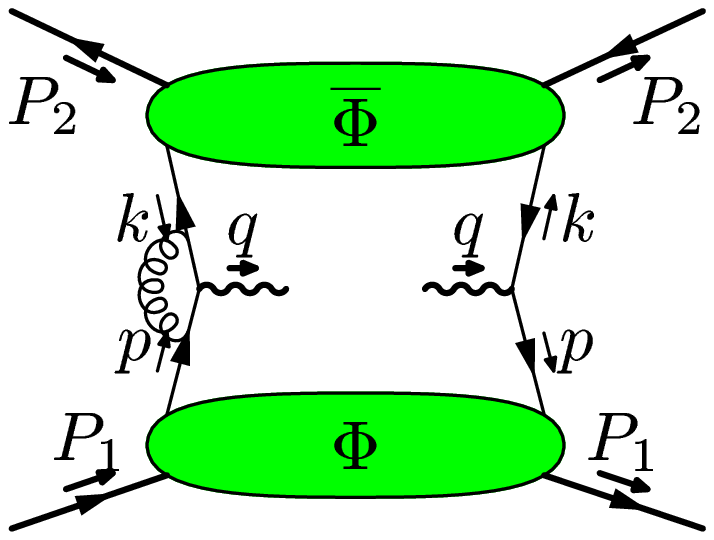}\\
& $\mathcal{L}^{[+]}$ &
  $\scriptstyle \frac{3}{8} \mathcal{L}^{[+]}{\rm Tr}\mathcal{L}^{[ \Box ]} -
    \frac{1}{8} \mathcal{L}^{[-]}$ &
  $\mathcal{L}^{[-]}$ &
  $\mathcal{L}^{[-]}$
\end{tabular}
\caption{The gauge links ($\mathcal{L}$) presented here are obtained as follows:
(1) start with a certain hard scattering diagram with correlators containing
no links, (2) sum over all diagrams of longitudinally polarized gluons
connecting a considered soft blob and the hard diagram in
the eikonal approximation. Gluons coupling
directly to the quark of the considered correlator are already present
in the soft blob definition, (3) the sum results in a link in the considered
soft blob multiplied by the \emph{same} hard scattering diagram.
\newline
A: various gauge links.
B: tree-level SIDIS and DY, cross-sections and leading contributions
in parton model with gauge links for $\Phi$.
C: corrections to SIDIS and DY
with gauge links for lower blobs; the coupling of the longitudinally polarized
gluons to the explicitly drawn gluon
\emph{has to be included} to obtain
a proper gauge link; the gauge links for the virtual corrections
have only be verified to the first non-trivial order.\label{links}}
\vspace{-.3cm}
\end{center}
\end{figure}

Gauge links for hard scattering diagrams can be derived
by coupling on-shell
longitudinally polarized gluons to the diagrams.
The delicate use of Ward identities in these calculations
is illustrated by the following QED
example. When considering S-matrix elements the sum of
coupling a photon with $\epsilon = p_1$ (and momentum $p_1\sim n_+$)
to all possible places in the tree-level amplitude
$\gamma^*(q-p_1) + e(k) \rightarrow e(p)$
vanishes, giving the relation
\begin{equation}
\parbox{2cm}{\includegraphics[width=1.8cm]{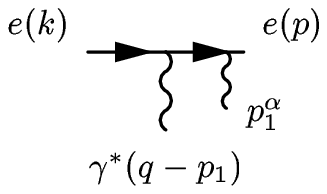}}
 =\ -
\parbox{2cm}{\includegraphics[width=1.8cm]{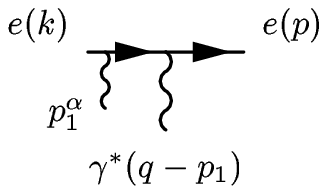}}.
\label{ward1}
\end{equation}
However, in link derivations
where the photon polarization
is the momentum \emph{direction}, the sum does not vanish. In fact, the sum
equals a gluonic pole matrix element\cite{Boer,Bomhof} which
could produce a SSA by itself
\begin{equation}
\parbox{2cm}{\includegraphics[width=1.8cm]{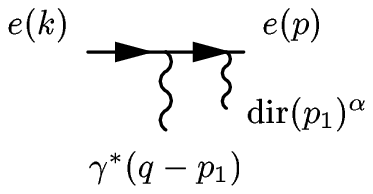}}+
\parbox{2cm}{\includegraphics[width=1.8cm]{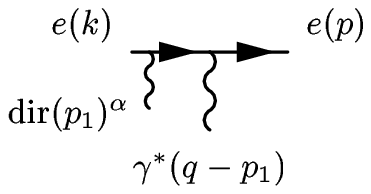}} =
\parbox{2cm}{\includegraphics[width=1.8cm]{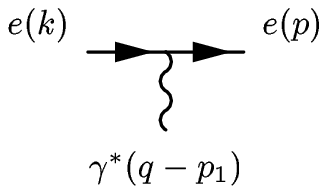}}
 \big( \frac{1}{p_1^+ + i\epsilon} +
\frac{1}{p_1^+ - i \epsilon} \big).
\end{equation}
Since similar effects also appear in QCD we refrained
from using identities like Eq.(\ref{ward1}).
Summing over the gluons explicitly
the link is straightforwardly derived to all orders
in the coupling and allows
for easy consistency checks by others
(are the first orders of the link correct and
is the quark-quark correlator gauge invariant?).
Some obtained results\cite{Boer,Bomhof,Pijlman} are given in
Fig.\ref{links}.

Gauge links form an
essential ingredient in considerations on factorization. In
factorization one typically tries to absorb gluon radiation with small
transverse momentum in a soft blob hoping that the constructed soft
blobs are in some sense universal.
From Fig.\ref{links}C it becomes clear that
the behaviour of gauge links when gluons are radiated depends on the process.
Although still calculable, the radiated gluon in DY in
Fig.\ref{links}C2 needs to be absorbed when constructing
the upper blob, but since the
gluon affects the gauge link of the lower blob, it will be
difficult - if not impossible - to factorize such
diagrams\cite{Pijlman}. This result appears beyond the
explicitly considered one-loop calculations
of the earlier discussed references.

Similar effects appear in other hadron-hadron scattering processes and
in fragmentation functions
in SIDIS. They do not appear in distribution
functions in
lepton and photon-hadron scattering, and
in fragmentation functions in $e^+ e^-$ annihilation.

\section{Experimental and theoretical input}

We would like to stress
that in those processes where one is not sensitive to the transverse
momenta of the quarks, one is dealing with
transverse momentum integrated
distribution and fragmentation functions. The links connect the
two quark fields by a straight line and are process independent.
Therefore, integrated SIDIS and
DY have no problems regarding their link structures.
As such, transversity can be best accessed via integrated DY,
$\Lambda$ polarization in SIDIS or two hadron fragmentation in
SIDIS\cite{BacchettaRadici}.

To understand
transverse momentum dependent
factorization and universality we need
experimental and theoretical input.
A comparison of a T-odd distribution function, such as the Sivers function,
in SIDIS and DY can have the following outcomes:
(1) they only differ by a sign and
the processes apparently factorize and links have predicting power,
(2) they are both zero (for some unknown symmetry),
a factorization theorem for DY probably exists
and the Lorentz invariance relations might hold, (3)
they are totally different and factorization, as we understand it now, is
violated.

It has been advocated\cite{CollinsMetz,Metz} that
fragmentation functions are independent of the link direction. 
Although the supporting
model\cite{Metz} ignores non-perturbative quarks
in the nucleon with $p^2\! > \! m^2$ which could
be sensitive to the link direction, the scenario itself remains possible.
If fragmentation shows up to be link independent,
then factorization in SIDIS
is probably feasible. Extended models
or experimental evidence (compare $z$ dependences of 
$D_1(z,P_h^{\perp2})$ \enlargethispage{\baselineskip}
or $H_1^{\perp(1)}(z)$ of SIDIS with $e^+ e^-$)
would contribute\ \nolinebreak a\ \nolinebreak lot.

This work, done in collaboration with C. Bomhof and P. Mulders,
was presented at a nicely organized conference SPIN 2004 in Trieste.
D. Boer,
L. Gamberg, P. H\"agler and A. Metz are acknowledged for useful discussions.

\end{document}